\documentclass[prb,
superscriptaddress,showpacs,amsmath,amssymb]{revtex4}

\usepackage{color}

\begin{document}

\title{From gravastar to central singularity}

\author{G.E.~Volovik}
\affiliation{Landau Institute for Theoretical Physics, acad. Semyonov av., 1a, 142432,
Chernogolovka, Russia}

\date{\today}

\begin{abstract}
We consider the model of the regular black hole, which demonstrates that the gravastar is thermodynamically unstable towards the  Schwarzschild black hole with singularity.
\end{abstract}
\pacs{
}

\maketitle
\tableofcontents

\section{Introduction}

Gravastar is the popular topic of discussions.\cite{Mottola2023,Visser2023,Mazur2004,Dymnikova1992} Gravastar is a compact, non-singular solution of Einstein’s equations -- the alternative to the Schwarzschild black hole with central singularity. It has an interior de Sitter region and an exterior Schwarzschild geometry,  which are separated by a phase boundary. This horizonless object has no central singularity. Its entropy is determined by the thermodynamics of the thin shell and it  is much smaller than the entropy of the Schwarzschild black hole. 

There is however another view,\cite{Ovalle2025,Kamenshchik2026,Kamenshchik2026b,Aoki2026,Lobo2026,Lobo2026a}
that the interior of the black hole can undergo a transition from being regular to containing a central singularity of the integrable type, see also \cite{Maeda2025}.

To resolve this controversy  we consider a simplified model of gravastar, in which the surface energy and its entropy are neglected. This model demonstrates that the gravastar is thermodynamically unstable towards the  Schwarzschild black hole with singularity, i.e. the process of the transition is accompanied by increasing entropy. The particular scenarios of the decay is via the de Sitter bubble inside the black hole. The bubble contracts and its entropy increases. Finally it forms the small size core with Planck energy density -- the Planck star.\cite{Rovelli2014} This is a kind of the singular core, which in the further step transforms to the central singularity.

\section{Gravastar with de Sitter core}

\subsection{Black and white holes}

We shall use Painleve-Gullstrand (PG) metric form:\cite{Painleve,Gullstrand}   
\begin{equation}
ds^2= -dt^2 + (dr - v(r) dt)^2 +r^2 d\Omega^2\,,
\label{PG}
\end{equation}
where $v(r)$ is the shift velocity. For the black and white holes the shift velocities are opposite:
\begin{eqnarray}
v(r)= -\sqrt{\frac{R}{r}} \,\,,\,\, {\rm black\,\, hole} \,,
\label{BH}
\\
v(r)= \sqrt{\frac{R}{r}} \,\,,\,\, {\rm white \,\, hole}\,.
\label{WH}
\end{eqnarray}
Here $R=2M$ is the radius of the black or white hole and $M$ is its mass (here we use $G=\hbar=c=1$). The opposite sign of the shift vector leads to the negative value of the white hole entropy: 
\begin{equation}
S_{\rm WH}(M)=-S_{\rm BH}(M)=-4\pi M^2 \,.
\label{WHentropy}
\end{equation}
The Eq.(\ref{WHentropy}) can be obtained in different ways including the processes of quantum tunnelling in which the rate of the process is determined by the difference in the entropy between initial and final states.\cite{Volovik2021c,Volovik2021a,Wilczek2000}

Black and white holes represent two degenerate states of an object with mass $M$, which corresponds to spontaneously broken time reversal symmetry. However, this only applies to the energy of the states. The same broken time reversal symmetry leads to opposite behavior of their thermodynamics, with opposite signs of their entropies.

\subsection{Black-white gravastar}

In the PG form, the horizonless gravastar has the following structure:\cite{Volovik2024}
\begin{eqnarray}
v(r)= -\sqrt{\frac{R}{r}} \,\,,\,\, r >R \,,
\label{v1}
\\
v(r)=- \frac{r}{R} \,\,,\,\, r < R \,,
\label{v2}
\end{eqnarray}
The core of this black hole corresponds to the contracting de Sitter state $v(r)=Hr$ with negative Hubble parameter, $H=-1/R$. The horizonless gravastar has zero entropy, since the positive entropy $A/4G$  of black hole horizon and the negative entropy $-A/4G$ of "cosmological" horizon of the contracting de Sitter cancel each other.

For the white-hole gravastar, the shift vector has opposite direction, and  the negative entropy $-A/4G$  of the white hole horizon cancels the positive entropy $A/4G$ of the "cosmological" horizon of the expanding de Sitter. These two objects, the black-hole gravastar  and the white-hole gravastar, are actually the equivalent states. 
These objects have no horizon, and thus their metrics can be written in the conventional Schwarzschild form, which does not experience the coordinate singularity at the horizon:
\begin{equation}
ds^2= -\left(1-v^2(r) \right)dt^2 +\left(1-v^2(r)\right)^{-1}dr^2 + r^2 d\Omega^2\,.
\label{Schwarzschild}
\end{equation}
The states with positive and negative shift functions $v(r)$ are indistinguishable. The time-reversal symmetry is not broken, and the $Z_2$ degeneracy no longer exists.

\section{Black hole with contracting de Sitter bubble as intermediate state between gravastar and Schwarzschild black hole}

\subsection{Black hole with contracting de Sitter bubble}

As the intermediate state between the gravastar and Schwarzschild black hole we consider   the de Sitter bubble, which occupies only the part of the interior of the black hole. Let us consider such state in the limit in which the radius of de Sitter bubble is much smaller than the radius of the black hole horizon, $r_0\ll R$, but the bubble is still classical. This limit case corresponds to
\begin{equation}
\frac{E_{\rm P}^2}{M} \ll |H|\ll E_{\rm P}\,,
\label{SmallBubble}
\end{equation}
where $E_{\rm P}$ is the Planck energy.
In this limit,  instead of the central singularity one has the small de Sitter bubble with radius $r_0$ such that 
\begin{equation}
 r_0^3(H)=\frac{2M}{H^2}\ll R^3\,.
\label{BubbleRadius}
\end{equation}

The dS bubble contains the "cosmological horizon" with radius $r_H= 1/|H|$, which is much smaller than the radius of the bubble,  
\begin{equation}
r_H^3= 1/|H|^3 =\frac{1}{2M|H|}\,\frac{2M}{H^2} =\frac{r_0^3(H)}{2M|H|} \ll r_0^3(H)  \ll R^3\,.
\label{SmallBubble2}
\end{equation}

In this limit the bubble mimics the homogeneous de Sitter Universe, which size is much larger than the cosmological horizon, but the Universe is well inside the black hole horizon. 

\subsection{Thermodynamics of de Sitter bubble}

Since the radius of the Universe is much larger than the radius of its cosmological horizon, $r_0\gg r_H$, we can use the thermodynamics laws of  the homogeneous de Sitter Universe.
The local temperature of the homogeneous de Sitter state and its entropy density and energy density  are:\cite{Volovik2024}
\begin{equation}
 T_{\rm dS}=\frac{H}{\pi} \,\,, \,\, s_{\rm dS}=\frac{3}{4} H  \,\,, \,\,  \epsilon_{\rm dS}=\frac{3}{8\pi} H^2 \,.
\label{EntropyDensity}
\end{equation}
For the expanding de Sitter spacetime, the local temperature is twice the Gibbons-Hawking temperature $T_{\rm GH}=H/2\pi$, which is related to the cosmological horizon.
For the bubble with contracting de Sitter, where $H<0$, the temperature $T_{\rm dS}$ and entropy density $s_{\rm dS}$ are both negative.

Since the total energy of the dS bubble of volume $V=(4\pi/3)r_0^3$ is equal the black hole mass, $M=\epsilon_{\rm dS}V$, the entropy of the bubble is:
\begin{equation}
 S_{\rm dS}(H)= s_{\rm dS}V=  s_{\rm dS}  \frac{M}{ \epsilon_{\rm dS}} = - \frac{2\pi M}{|H|}\,.
\label{EntropyDS}
\end{equation}
The total entropy of the black hole with de Sitter bubble is the sum of the entropy of dS bubble and  the entropy of the black hole horizon  $S_{\rm BH}=4\pi M^2$:
\begin{eqnarray}
 S_{\rm BH}(M,H)=  S_{\rm dS}(H) + S_{\rm BH}(M) = 
\label{EntropyGr}
 \\
 =S_{\rm BH} (M)\left(1- \frac{1}{2M|H|} \right) = S_{\rm BH} (M)\left(1- \frac{1}{R|H|} \right).
\label{EntropyGr2}
\end{eqnarray}
At $|H| \rightarrow 1/R$, the entropy tends to zero, which is natural for the static gravastar. Since Eq.(\ref{EntropyGr2}) is valid in both limits, $|H| \gg 1/R$ and  $|H| \rightarrow 1/R$, this demonstrates that the suggested model is relevant for description of the  thermodynamics of the black hole with de Sitter bubble, and Eq.(\ref{EntropyGr2}) can be used for the qualitative consideration in the whole range of $|H| \geq \frac{1}{2M}$.

\subsection{On Bekenstein-Hawking entropy of black hole}

The sum of entropies in Eq. (\ref{EntropyGr}) can be interpreted in the following way. It is the sum of entropies of the regions $r<r_0$ and $r_0<r<\infty$. The entropy of the first region comes from the dark energy of the de Sitter bubble,  $S(r<r_0)=S_{\rm dS}(H)$. The region $r_0<r<\infty$ does not contain matter fields, but it contains the event horizon at $r=R$. The horizon, being the source of irreversibility, contributes the entropy $S(r>r_0)=S_{\rm hor}=\pi R^2=S_{\rm BH}(M)$.  

On the other hand, the horizon entropy can be also obtained from Eq.(\ref{EntropyGr}). When $|H|\rightarrow 1/R$, one obtains the horizonless gravastar, which has zero entropy, $S_{\rm BH}(M,H=-1/R) = 0$. Then from Eq.(\ref{EntropyGr}) one obtains
\begin{equation}
 S_{\rm BH}(M)=-S_{\rm dS}(H=-1/R)=\pi R^2=\frac{A}{4}\,.
\label{EntropyBH}
\end{equation}
This shows that the local thermodynamics of the de Sitter state in Eq.(\ref{EntropyDensity}) allows us to connect the entropy of the Schwarzschild black hole  $S_{\rm BH}(M)$, which depends on the mass inside the horizon,
 with the Bekenstein-Hawking entropy of the event horizon $A/4$, which depends on the area of the horizon. In a similar way, the entropy of the region inside the cosmological horizon  is equal to the entropy of the cosmological horizon, $(4\pi/3)r_h^3 \,s_{\rm dS}=A_h/4$. These are two examples of the holographic bulk-horizon correspondence.

The black hole entropy, which is interpreted as the entropy of event horizon, can be also interpreted in terms of the statistical ensemble of the black holes, which obeys the non-extensive Tsallis-Cirto statistics. The Tsallis-Cirto $\delta=2$  statistics\cite{TsallisCirto2013} governs the processes of splitting and merging of black holes:
\begin{equation}
  \sqrt{S_{\rm BH}(M)}= \sqrt{S_{\rm BH}(M_1)} +\sqrt{S_{\rm BH}(M_2)} \,\,,\,\, M=M_1 +M_2\,.
\label{CompositionBH}
\end{equation}
The probability of the splitting of the black hole into two smaller black holes by quantum fluctuations is $p \propto e^{\Delta S}$, where $\Delta S$ is the change in entropy in this process:\cite{Volovik2021c,Volovik2021a,Wilczek2000}
\begin{equation}
 \Delta S=S_{\rm final} - S_{\rm initial} = - 2 \sqrt{S_{\rm BH}(M_1)S_{\rm BH}(M_2)} \,.
\label{SplittingBH}
\end{equation}

In this interpretation, the entropy of a black hole determines the probability 
 of the complete annihilation of the black hole by quantum fluctuation, $p \propto  \exp(S_{\rm final} - S_{\rm initial})\propto \exp(-S_{\rm BH}(M))$.  Example of such fluctuation is the process of macroscopic quantum tunnelilng, in which the black hole splits into $N=M/E_{\rm P}$ elementary black holes with Planck mass -- Planckons.\cite{Volovik2025} Here  $S_{\rm final} \propto N$, which is small compared with $S_{\rm initial}=S_{\rm BH}(M) \sim N^2$.

But in fact, the entropy of a horizon is a property of a statistical ensemble of horizons, since  the non-extensive Tsallis-Cirto statistics with $\delta=2$ also regulates the composition law  for the entropies of the horizons inside the black hole. In particular, the composition law  for the entropies of the inner horizon at $r=r_-$ and the outer horizons at $r=r_+$ of the Reissner-Nordstr\"om black hole is:\cite{VolovikRN} 
\begin{equation}
\sqrt{S_{\rm RN}}= \sqrt{|S(r=r_-)|}+\sqrt{S(r=r_+)}\,.
\label{CompositionHor}
\end{equation}
This composition law combines the positive entropy of the outer horizon and the negative entropy  of the inner horizon. The entropy in the statistical ensemble of the horizons can provide an answer to the question posed by Unruh:\cite{Unruh2026} "Do Black Holes possess entropy or do they create it?".

\subsection{White hole with expanding de Sitter bubble}

 The regular white hole can be considered in the same manner as we considered the regular black hole. Now instead of the central singularity there is the de Sitter bubble with $H>0$ and thus with positive entropy. The total entropy of such anti-gravastar can be written as
 \begin{eqnarray}
 S_{\rm WH}(M,H)=S_{\rm WH}(M) + S_{\rm dS}(H)=
  \nonumber
 \\
= S_{\rm WH} (M)\left(1- \frac{1}{2MH} \right)=  - S_{\rm BH} (M)\left(1- \frac{1}{RH} \right).
\label{EntropyAntiGr}
\end{eqnarray}

At $H=1/R=1/(2M)$ one obtains again the static state with zero entropy, where the anti-gravastar is indistinguishable from the gravastar.

\subsection{Thermal instability of static gravastar}

From Eq.(\ref{EntropyGr2}) it folows that with decreasing $|H|$ towards $1/(2M)$, the de Sitter bubble grows until its volume $V$ reaches the volume inside the black hole horizon with $R=2M$. In this state, the negative entropy of dS bubble cancels the black hole horizon entropy, and  one obtains the static horizonless gravastar with zero entropy.  The same static gravastar with zero entropy is obtained when  $H$ decreases towards $1/(2M)$ in the white gravastar in Eq.(\ref{EntropyAntiGr}). The gravastars with $H=\pm 1/2M$ are indistinguishable. They represent the static gravastar, which consideration does not require the transition to the Painleve-Gullstrand coordinates.

On the way from the white hole to static gravastar, the entropy increases from $S_{\rm WH}=-S_{\rm BH}= - 4\pi M^2$ to zero. This demonstrates that the white hole is thermodynamically unstable towards the static gravastar. The entropy also increases on the way from zero in static 
gravastar to the black hole with entropy  $S_{\rm BH}= 4\pi M^2$.  This demonstrates that the static gravastar is unstable towards the black hole with singularity.

Note that we do not consider the real dynamical route from the regular gravastar to the black hole with central singularity. The consideration is pure thermodynamical, and the thermodynamic route does not coincide with the dynamic path, which is determined by equations of motion. In the thermodynamic route, entropy increases from almost zero gravastar entropy to the entropy of the Schwarzschild black hole. We do not take into account the gravastar boundary. The boundary is important for the existence of the gravastar, but its entropy is negligibly small compared to the entropy of a Schwarzschild black hole, and thus it can be neglected in thermodynamics. Of course, in a dynamic context, the boundary may create an energy barrier along this path; however, this does not alter the fundamental tendency toward increasing entropy. The barrier will finally be overcome, say, via quantum tunneling.

The instability of gravastar can be triggered by the same instability toward the creation of the pairs of particles in the de Sitter state in general.\cite{Volovik2026proton} In the cosmological de Sitter phase, the de Sitter state is homogeneous, i.e. all the points of the de Sitter vacuum are equivalent, and the local temperature is the same for all points in the de Sitter space. Under this condition, the pair creation can be realized only if some external object, such as proton, violates this translation symmetry. 
 The rate of the pair creation by external object is $w\propto \exp(-\pi\epsilon/H)$, where $\epsilon$ is the energy of the created pair. This is also applicable to the process of decay of the bound states in de Sitter environment, such as ionization of the hydrogen atom, where $\epsilon$ is the activation energy, see e.g.\cite{Maxfield2022}. Both processes are described by the same local activation temperature, which is twice the Gibbons-Hawking temperature.  The created particles in turn  become the source of the creation of the other particles, etc. As a result the energy of the created matter leads to decrease of the vacuum energy. 

In case of the de Sitter core in the gravastar, the translational symmetry is violated, and the pair creation becomes possible even without the external object. But now we have de Sitter vacuum with $H<0$.
Similar to the white hole and to the inner horizon of the RN black hole, such state is described by negative temperature.  The negative temperature and the corresponding negative entropy are supported by many different considerations, see also Sec. II.A and references therein. In the gravastar case, the reverse sign means the change of the direction of the particle trajectories to the opposite. Instead of direction away from the centre at $H>0$, the trajectories at $H<0$ are directed toward the centre, where the created particles form the central singularity at $r=0$. Of course, one cannot describe the full dynamics of the matter creation and its further absorption by the forming singularity, but the thermodynamic tendency to increasing entropy should be valid. Within the framework of the thermodynamic approach, the created matter is absorbed by the singularity, leaving behind a de Sitter core with a steadily decreasing energy of the de Sitter bubble and a steadily increasing entropy until the de Sitter bubble fully shrinks.

\section{From dS bubble to the Planck-density core} 

\subsection{Planck-density core} 

The process of decay of gravastar towards the singularity leads to formation of the peculiar object -- the bubble with energy density of Planck scale, $\epsilon \sim E_{\rm P}^4$. This object, which is the precursor of the singularity, is formed when $|H|$ reaches the Planck energy scale, 
$|H| \sim E_{\rm P}$, and curvature reaches Planck level, ${\cal R}\sim 1/l^2_{\rm P}$. Then from Eq.(\ref{BubbleRadius}) one obtains the size of the  formed core:
  \begin{equation}
 r_0(E_{\rm P}) \sim  l_{\rm P}\, \left(\frac{M}{E_{\rm P}}\right)^{1/3} \gg l_{\rm P}\,.
\label{PlanckCore}
\end{equation}
The same result is obtained considering the Kretschmann scalar,\cite{Shaya2026} which for Schwarzschild black hole of mass $M$ is $K=48 M^2/r^6$. It becomes on the order of Planck scale for $r \sim r_0(E_{\rm P})$ in Eq.(\ref{PlanckCore}).

The size of the Planck-density core depends on the black hole mass $M$ and thus it can be macroscopically large, although it is much smaller the black hole radius. This macroscopic core with Planck energy density is a kind of the regular singularity (or integrable singularity\cite{Kamenshchik2026}). The entropy $S_{\rm core}$ of the Planck-density core is large, but still is much smaller than the black hole entropy, which is determined by the horizon:
 \begin{equation}
1 \ll S_{\rm core} \sim \frac{M}{E_{\rm P}} \ll S_{\rm BH}\sim \frac{M^2}{E^2_{\rm P}}\,.
\label{PlanckCoreEntropy}
\end{equation}

\subsection{From Planck-density core to central singularity} 

From Eq.(\ref{EntropyDensity}) it follows that the temperature of the Planck-density core is $T_{\rm core } \sim -E_{\rm P}$, which actually corresponds to  $T=-\infty$ or to inverse temperature $\beta=0$. The inverse temperature $\beta=0$ is the intermediate temperature between the negative and positive temperatures. It takes place on the way from the negative temperature  of the contracting de Sitter to the positive temperature $T_{\rm BH}=1/4\pi R$ of the "equilibrium" state of the black hole with singularity at $r=0$. At the central singularity, the spacetime ceases to be locally Minkowskian, which is called a "Minkowski breaking" \cite{Kamenshchik2026b}.

Another way for a de Sitter bubble to contract toward a singularity is to decrease its radius while maintaining a constant Hubble parameter $H$. This contraction is accompanied by the formation and continuous growth of singularity at $r = 0$. We did not consider the role of the singularity at the surface of the bubble at $r=r_0$, where the $dv/dr$ has jump. But in the Planck limit the surface energy is certainly much smaller than the bulk energy, and the same is in the limit $r_0\rightarrow R$, when the energy of the surface shell can be neglected.

\section{Round-the-world trip of black hole}

The black hole is the mixed state -- the thermodynamic state. According to Landau and Lifshitz, \cite{LL5}  such states experience the random fluctuations which decrease the entropy, and then the entropy is restored in the continuous thermodynamic process. Let us consider an example, in which the black hole experiences quantum fluctuation, which transforms the black hole to the white hole with the same mass $M$, and then the black hole is restored in the processes which include formation of the Sitter bubbles and gravastar.

Below is the particular route of the round-the-world trip of black hole from formation of the white hole by macroscopic quantum tunnelling to restoration of the black hole:

1) {\it White hole phase}. White hole is formed from the black hole  by macroscopic quantum tunnelling\cite{Volovik2021a}  -- the quantum fluctuation. In the same manner as the thermal fluctuations, this random fluctuation reduces the black hole entropy from its maximal positive value $S_{\rm BH}(M)=A/4G$
to its maximal negative value $S_{\rm WH}(M)=-S_{\rm BH}(M)=-A/4G$. The further thermodynamic processes step by step restore the original black hole entropy $S_{\rm BH}(M)=A/4G$.

2)  {\it White hole with de Sitter bubble}. Being the state with maximal negative entropy, the white hole is thermodynamically unstable towards  formation of the de Sitter bubble inside the white hole. This bubble contains the expanding de Sitter, which has the positive Hubble parameter, $H>0$, and positive entropy.  The de Sitter bubble grows, its entropy increases, and therefore the whole entropy of the white hole also increases.

3) {\it Static gravastar phase}. The boundary of the de Sitter bubble finally reaches the white hole horizon, and its positive entropy compensates the negative entropy of white hole horizon. This horizonless object with zero entropy  is invariant under time reversal and is equivalent to the static black-hole gravastar. 

4) {\it Black hole with de Sitter bubble}. The static gravastar with de Sitter core is thermodynamically unstable and transforms to black hole with the contracting de Sitter bubble inside the horizon. The negative entropy of the bubble does not compensate the positive horizon entropy, and the black hole entropy becomes positive. 

5) {\it Black hole phase with Planck-density core}. The de Sitter core continuously shrinks with increasing entropy and finally transforms to the Planck-density core with negligible entropy, so that the  the total entropy of the black hole is close to $A/4G$, but still less.

6) {\it Schwarzschild black hole}. Finally the Planck-density core transforms to the central singularity of the Schwarzschild black hole -- the state of the maximal entropy $S_{\rm BH}(M)=A/4G$. 

There is a scenario in which our Universe is the inside of a black hole in a parent Universe, see e.g. Refs.\cite{Pathria1972,Brandenberger2001,Chakrabarty2001,Shamir2025} and references therein. If we draw a similar analogy between the de Sitter bubble and our Universe, it turns out that our Universe with its expanding de Sitter is in the stage 2 of its development, that is it is located inside a giant white hole, the radius of which is much greater than the radius of the cosmological horizon, $R \gg 1/H$ (see Eq.(\ref{SmallBubble2})). In this white hole scenario, our Universe would go through further stages of development (from 2 to 6 with increasing total entropy) and eventually collapse into the central singularity of a giant black hole.

\section{Conclusion}

We considered the oversimplified model of regular black hole. We did not take into account the surface tension of the boundaries between different regions of the black hole. Also the deformation of the vacuum by gravitational field inside the horizon (due to the presence of the negative energy states of fermions)\cite{Volovik2021,Selch2023} is not taken into account. Nevertheless the model clearly demonstrates that the gravastar with the de Sitter core is thermodynamically unstable towards the  Schwarzschild black hole with central singularity. It provides the route from the gravastar to  Schwarzschild black hole, in which the entropy continuously increases from the almost zero value to the black hole entropy $S_{\rm BH}=A/4G$. The thermodynamic arrow of time indicates that the Schwarzschild state with its central singularity is the equilibrium state of a black hole.

{\bf Acknowledgement.}
I thank Francisco Lobo for discussion, and Hideki Maeda and Jorge Ovalle for correspondence.


\begin{thebibliography}{999}

\bibitem{Mottola2023}
E. Mottola,
Gravitational Vacuum Condensate Stars,
Chapter in the book:
{\it Regular Black Holes, Towards a New Paradigm of Gravitational Collapse},
Editor: Cosimo Bambi, Springer, Pages 283--352,
arXiv:2302.09690 [gr-qc].

\bibitem{Visser2023}
R. Carballo-Rubio, F. Di Filippo, S. Liberati and M. Visser,
Singularity-free gravitational collapse: From regular black holes to horizonless objects,
Chapter in the book:
{\it Regular Black Holes, Towards a New Paradigm of Gravitational Collapse},
Editor: Cosimo Bambi, Springer, Pages 353--387,
arXiv:2302.00028 [gr-qc].

\bibitem{Mazur2004}
P.O. Mazur and E. Mottola,
Gravitational vacuum condensate stars,
PNAS {\bf 101}, 9545--9550 (2004).

\bibitem{Dymnikova1992}
I. Dymnikova, 
Vacuum nonsingular black hole,
Gen. Rel. Grav. {\bf 24}, 235--242 (1992). 

\bibitem{Ovalle2025}
J. Ovalle,
Interior evolution of Regular Schwarzschild Black Holes,
Phys. Lett. B {\bf 881}, 140912 (2026),
arXiv:2509.00816.

\bibitem{Kamenshchik2026}
R. Casadio, A. Giusti, A. Kamenshchik and J. Ovalle,
On gravitational collapse and integrable singularities,
 The European Physical Journal C {\bf 86}, 1018 (2026),
arXiv:2605.01808.

\bibitem{Kamenshchik2026b}
J. Ovalle, R. Casadio, and A. Kamenshchik,
Schwarzschild black hole singularity formation,
Phys. Rev. D {\bf 113}, 064042 (2026),
arXiv:2603.06451 [gr-qc].

\bibitem{Aoki2026}
S. Aoki and J. Ovalle, 
An analytic model for a total process of gravitational collapse: from star to Schwarzschild black hole, Eur. Phys. J. C {\bf 86}, 741 (2026). 

\bibitem{Lobo2026}
F.S.N. Lobo and M.E. Rodrigues,
Entropy release from Minkowski breaking in regular Schwarzschild black holes,
arXiv:2607.14079.

\bibitem{Lobo2026a}
F.S.N. Lobo and M.E. Rodrigues,
Entropy dynamics in gravitational collapse: From Minkowski breaking to de Sitter thermodynamics,
arXiv:2607.24349.

\bibitem{Maeda2025}
H. Maeda,
Fake Schwarzschild and Kerr black holes,
Class. Quant. Grav. 42 (2025) 23, 235005.
arXiv:2410.11937 [gr-qc].

\bibitem{Rovelli2014}
C. Rovelli and F. Vidotto,
Planck stars, Physics {\bf 23}, 12 (2014),
arXiv:1401.6562

\bibitem{Painleve} 
P. Painlev\'e, 
La m\'ecanique classique et la th\'eorie de la relativit\'e, 
 C. R. Acad. Sci. (Paris) {\bf 173}, 677 (1921).
 
 \bibitem{Gullstrand} 
A. Gullstrand,
Allgemeine L\"osung des statischen Eink\"orper-problems in der Einsteinschen Gravitations-theorie,
Arkiv. Mat. Astron. Fys. {\bf 16}, 1-15 (1922).

\bibitem{Volovik2021c}
G.E. Volovik,
From black hole to white hole via the intermediate static state,
Modern Physics Letters A {\bf 36}, 2150117  (2021),
arXiv:2103.10954.

 \bibitem{Volovik2021a}
G.E. Volovik,
Macroscopic quantum tunneling: from quantum vortices to black holes and Universe,
ZhETF {\bf 162}, 449--454 (2022), 
JETP {\bf 135}, 388--408 (2022),
arXiv:2108.00419.

\bibitem{Wilczek2000}
M.K. Parikh and F. Wilczek, 
Hawking radiation as tunneling,
Phys. Rev. Lett. {\bf 85}, 5042 (2000).

\bibitem{Volovik2024}
G.E. Volovik, 
Thermodynamics and decay of de Sitter vacuum,
Symmetry {\bf 16}, 763 (2024),
https://doi.org/10.3390/sym16060763

\bibitem{TsallisCirto2013}
C. Tsallis and L.J.L. Cirto,
Black hole thermodynamical entropy,
Eur. Phys. J. C {\bf 73}, 2487 (2013).


\bibitem{Volovik2025}
G.E. Volovik, 
Thermodynamics of black and white holes in ensemble of Planckons,
JETP {\bf 141}, 32--40 (2025),
%DOI: 10.1134/S1063776125601041
arXiv:2506.13145 [gr-qc].

\bibitem{VolovikRN}
G.E. Volovik, 
Thermodynamics of Kerr black hole: Tsallis-Cirto composition law and entropy quantization,
JETP Letters {\bf 123}, 509--513 (2026),
arXiv:2509.00748.

\bibitem{Unruh2026}
W.G. Unruh,
Black Hole -- Entropy Container or Creator using a simple Black Hole model and a simple linear amplifier model,
arXiv:2603.18374.

\bibitem{Volovik2026proton}
G.E. Volovik, 
Proton decay in de Sitter environment,
tbp in: Open Issues in Gravitation and Cosmology,
"Original Contributions, Essays, and Recollections in Honor of Alexei Starobinsky,"
Editors: Andrei Barvinsky, Alexander Kamenshchik, 
Part of the book series: Fundamental Theories of Physics, FTPH {\bf 225}, Springer 2026,
arXiv:2411.01892.

\bibitem{Maxfield2022}
H. Maxfield and Z. Zahraee, Holographic solar systems and hydrogen atoms: non-relativistic physics in
AdS and its CFT dual, JHEP 11 (2022) 093.


\bibitem{Shaya2026}
E.J. Shaya,
The Quantum Boundary of Black Hole Interiors: Path-Integral Termination at Planck Curvature,
arXiv:2606.15423.

\bibitem{LL5}
 L.D. Landau  and  E.M. Lifshitz, 
 Course of Theoretical Physics, Volume 5, Statistical Physics, Elsevier, 1980.
 
  \bibitem{Pathria1972}
 R.K. Pathria,
 The Universe as a Black Hole,
  Nature {\bf 240}, 298 (1972).
 
 \bibitem{Brandenberger2001}
D.A. Easson and R.H. Brandenberger,
Universe Generation from Black Hole Interiors,
JHEP 0106:024,2001,
arXiv:hep-th/0103019.

 \bibitem{Chakrabarty2001}
H. Chakrabarty, A. Abdujabbarov, D. Malafarina and C. Bambi,
A toy model for a baby universe inside a black hole,
Eur. Phys. J. C {\bf 80}, 373 (2020).

 \bibitem{Shamir2025}
 L. Shamir,
The distribution of galaxy rotation in {\it JWST} Advanced Deep Extragalactic Survey,
 MNRAS {\bf 538}, 76--91 (2025).


 \bibitem{Volovik2021}
G.E. Volovik. 
Type-II Weyl Semimetal versus Gravastar,
 JETP Lett. {\bf 114}, 236--242 (2021),
 arXiv:2106.08954.

\bibitem{Selch2023}
M. Selch, J. Miller and M.A. Zubkov,
Gravastar-like black hole solutions in $q$-theory,
Class. Quantum Grav. {\bf 40}, 155017 (2023),
arXiv:2301.02914.

\end{thebibliography}
\end{document}